\documentclass[12pt]{iopart}

\usepackage{graphicx}
\usepackage{amssymb}

\def\beeq{\begin{equation}}
\def\eneq{\end{equation}}
\def\beeqa{\begin{eqnarray}}
\def\eneqa{\end{eqnarray}}

\begin{document}

\jl{3} 

\title{Liquid-liquid phase transition in  Stillinger-Weber silicon}
 
\author{Philippe Beaucage and Normand Mousseau \footnote{Normand.Mousseau@umontreal.ca}} 
\address{D\'epartement de physique and Regroupement qu\'eb\'ecois
   sur les mat\'eriaux de pointe, Universit\'e de Montr\'eal, C.P.
   6128, succ. Centre-ville, Montr\'eal (Qu\'ebec) H3C 3J7, Canada}

\date{\today}

\begin{abstract}
  
  It was recently demonstrated that the Stillinger-Weber silicon
  undergoes a liquid-liquid first-order phase transition deep into the
  supercooled region (Sastry and Angell, Nature Materials 2, 739
  (2003)). Here we study the effects of perturbations on this phase
  transition.  We show that the order of the liquid-liquid transition
  changes with negative pressure. We also find that the liquid-liquid
  transition disappears when the three-body term of the potential is
  strengthened by as little as 5 \%. This implies that the details of
  the potential could affect strongly the nature and even the
  existence of the liquid-liquid phase.

\end{abstract}

\pacs{
64.70.Ja   
61.20.Ja   
61.43.Dq  
 }

\maketitle
\section{Introduction}
\label{intro}

The amorphous phase of Si is particular in that it does not correspond
to the arrested liquid phase, contrarily to glasses: while liquid Si
is metallic, with an average coordination around 6.4 at ambient
conditions~\cite{Waseda3, Waseda, Egry}, {\it a}-Si has a coordination
near 4 and is a semiconductor.  The existence of a possible
intermediate phase, explaining in part this difference, was first
suggested by Aptekar, who showed that the Gibbs free energy of the
amorphous phase does not extrapolate smoothly to that of the liquid,
indicating that an additional phase transition should occur at around
1450 K ~\cite{Aptekar}. Much experimental~\cite{Ansell, Donovan,
  Thompson2, Baeri} and numerical work~\cite{Angell2, Angell,
  Luedtke2, Luedtke} followed, supporting the existence of such an
additional phase, long thought to be the amorphous phase.

A breakthrough in the understanding of this unusual feature came a few
years ago with the first clear experimental evidence for liquid
polymorphism in a number of materials such as
Y$_2$O$_2$--Al$_2$O$_3$~\cite{aasland}.  The distinct liquid-phases
hypothesis was first formulated to explain several properties of water
near the melting temperature, including the density anomalies, and was
supported by the polymorphism of the crystal and amorphous phases.
More precisely, Mishima {\it et al.}~\cite{Mishima1, Mishima2}
proposed that the liquid-liquid transition for water between a low
density liquid (LDL) and a high density liquid (HDL) is the continuity
at higher temperatures of the known amorphous phase transition between
the low and high density forms.

In analogy with water, it was rapidly suggested that a
``fragile-to-strong'' liquid transition occurs in the supercooled
regime of silicon, the resulting viscous liquid at low temperature
(i.e. LDL) corresponding to the precursor phase of the amorphous
metastable state~\cite{Angell4,Angell3, Angell}; the best candidates
for such a liquid-liquid phase transition are tetravalent systems,
which display an open molecular structure, such as H$_2$O ,C, Si, Ge,
SiO$_2$ and GeO$_2$ which all show a density maximum in function of
temperature~\cite{Poole1, Sciortino, Saika}.  {\it Ab initio}
numerical calculations \cite{Durandurdu} also show a transition
between a low (LDA) and high (HDA) density amorphous phase of silicon
under high pressure, providing an additional support for this
hypothesis.  More importantly, the liquid-liquid transition for
silicon is not excluded from the stable region of the phase diagram
and recent experimental work on liquids revealed a structural change
for liquid Si \cite{Funamori}, GeO$_2$ ~\cite{Ohtaka} and P
~\cite{Katayama} under high pressure.

Recently, the first clear demonstration of the existence of this
low-density liquid phase, at least for the Stillinger-Weber silicon,
was given by Sastry and Angell~\cite{Sastry}.  In this elegant work, Sastry
and Angell identify a first-order liquid-liquid phase transition
taking place around 1060 K by measure the heat release in a simulation
in the NPH ensemble, finding results in agreement with a number of previous
simulations. They showed, moreover, that the resulting phase is a very
viscous tetrahedral liquid which should be the precursor to the
well-characterized amorphous phase.

Although this pioneering work establishes the existence of the
liquid-liquid phase transition, it is essential to characterize the
behavior of this transition under perturbations.  In this Paper, we
study the impact of pressure and potential modifications on this
transition. In particular, we find that the amorphous phase does
correspond to the glassy low-density liquid at zero pressure.
However, we also find that the liquid-liquid transition changes order
with negative pressure. We show that this transition becomes
unreachable with a very slight change of the SW potential, indicating
that while this transition is clearly present for this potential, it
remains to be fully demonstrated for real Si.

\section{Methods} \label{meth}

The molecular dynamical simulations for this work are performed in
three ensembles: isobaric (NPE), isothermal-isobaric (NPT) and
iso-enthalpic-isobaric (NPH). All simulations are done at P=0 in a
cubic box containing 1000 atoms, with periodic-boundary conditions.
The extended-system method of Andersen is used to control
pressure~\cite{Andersen, Haile, Brown} and Hoover's constraint method
for the temperature~\cite{Hoover, Evans, Allen}.  We use the
fifth-order Gear predictor-corrector to integrate Newton's equation
with a time step $\Delta t = 1.15$ fs.  Typically, after a change in
temperature, the simulations are equilibrated for 50 000 $\Delta t$
(58 ps) and statistics are accumulated over 450 000 $\Delta t$ (518
ps).

Atomic interactions are represented by the Stillinger-Weber potential
(SW), developed to reproduce accurately the crystalline and the liquid
state of Si~\cite{SW85}. Although this potential has known
limitations, especially in the amorphous phase
~\cite{Luedtke,Luedtke2,Barkema96, Vink}, it ensures a reasonable
description of the liquid phase ~\cite{Cook, Balamane, Broughton}.
Its melting temperature is also near the experimental value of 1683 K
~\cite{Ashcroft}: averaging temperature of a box with a crystal-liquid
interface at equilibrium, we find a temperature of 1662 $\pm$ 1 K, in
agreement with $T_m=1665$ K found by Landman {\it et
  al.}~\cite{Landman3} using a similar method.

\begin{figure}
\centering
\begin{tabular}{ccc}
\includegraphics[width=0.9in]{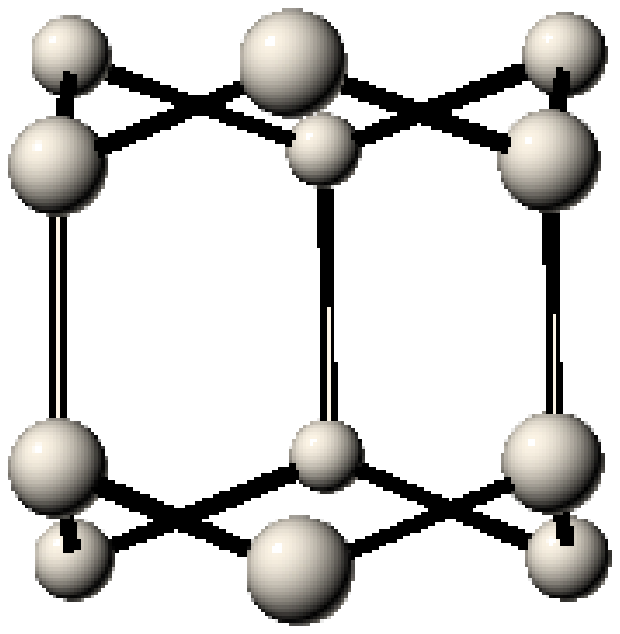} & 
\includegraphics[width=1.0in]{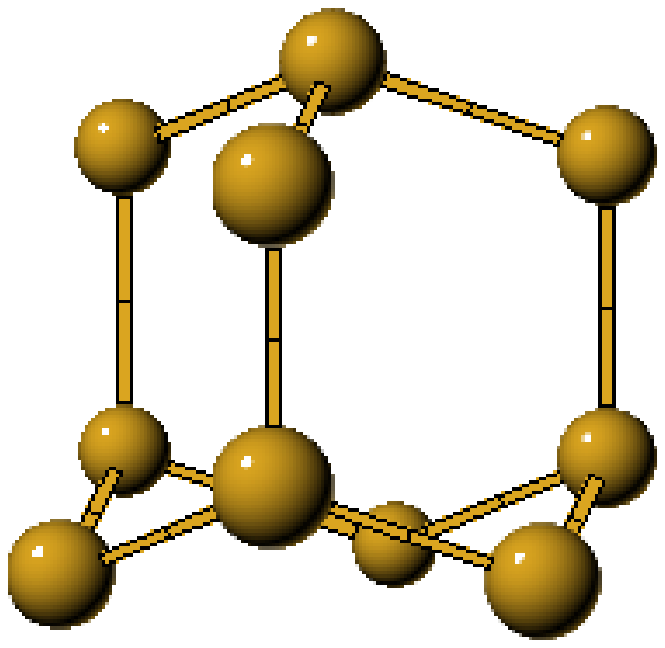} &
\includegraphics[width=1.0in]{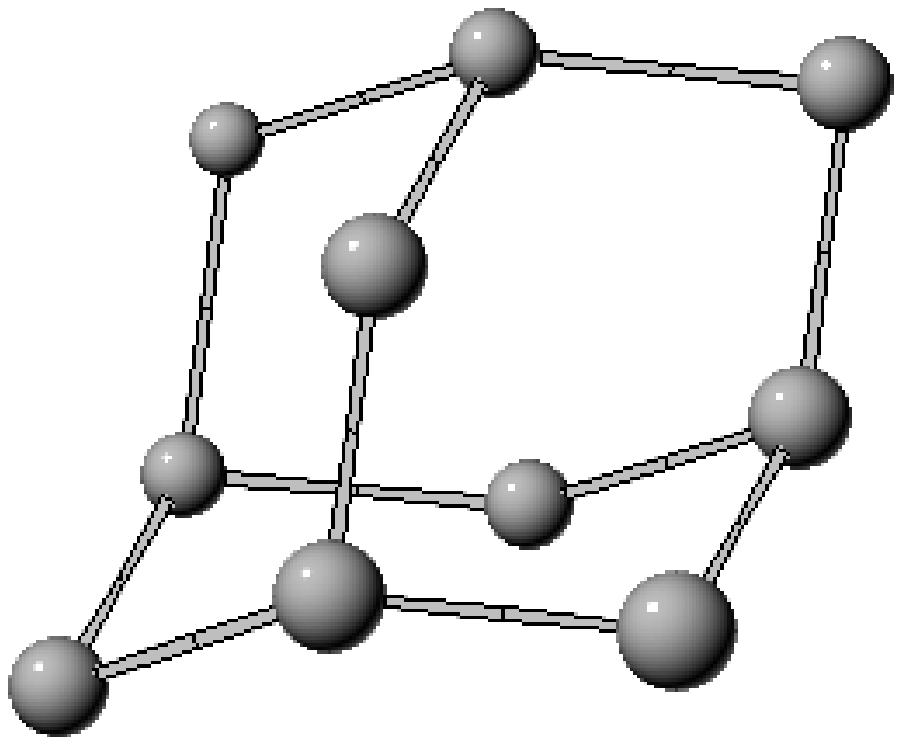} 
\end{tabular}
\caption{The three basic building blocks which
  represent the order parameter for characterizing the state of
  crystallization of the model.  The wurtzite basic block (left) is a
  12-atom cluster composed of two sixfold rings whereas the diamond
  basic block (center) is a 10-atom cluster with four sixfold rings.
  The $\beta$-tin basic block (right) is equal to a diamond basic
  block where the tetrahedra are compressed in one direction and
  elongated along the two others axes.}
\label{basic_b}
\end{figure}

To measure directly the degree of local crystallinity in the liquid
and amorphous phases, we use a set of criteria that identify the
smallest three-dimensional structures associated with wurtzite,
diamond and $\beta$-tin crystalline structures (see Fig.
~\ref{basic_b}). We restrict this topological analysis to pairs of
atoms that are within 2.75 \AA\, of each other to ensure that these
atoms are close to a crystalline or amorphous environment; for all
others quantities, we consider the first-neighbor cut-off to be at the
first minimum between the first- and second-neighbor peaks in the
radial distribution function (RDF).  The wurtzite elementary block is
a 12-atom cluster with two six-fold rings stacked on top of each other
and connected by three bonds; the elementary building block for both
diamond and $\beta$-tin has the same topology consisting of four
six-membered rings placed back to back, forming a 10-atom cluster.
These elementary clusters are only present with a low density in
good quality amorphous Si relaxed with SW (10 to 20 at.\%) as well as
in the HDL (5-10 at.\%) providing a very convenient measure of local
crystalline order; they were used for the same purpose in a previous
study of crystallization ~\cite{Nak}. This order parameter is a much
more sensitive measure of crystallinity than the structure factor or
the RDF.

\section{Results}
\label{resul}

\subsection{Liquid-liquid transition}
\label{trans_l-l}

\subsubsection{Zero pressure}
\label{P_zero}

The Stillinger-Weber liquid-to-amorphous transition was one of the
first problems studied after the potential was introduced. Broughton
and Li observed that supercooled silicon transforms continuously into
a glass at low temperature~\cite{Broughton}. Further investigations by
Luedtke and Landman, however, showed that the liquid undergoes a
first-order phase transition at around 1060 K, transforming directly
into a reasonnably good quality amorphous solid if the simulated
cooling down is slow enough~\cite{Luedtke, Luedtke2}. This phase was
found to be reversible as the system is heated from the amorphous
phase.  Following recent work on polymorphism in liquids, Angell and
Borick~\cite{Angell2} suggested that the phase just below the
transition temperature is not an amorphous solid but a very viscous
liquid, freezing into the amorphous phase. As discussed in the
introduction, this suggestion was recently demonstrated by Sastry and
Angell using an elegant analysis~\cite{Sastry}: simulating
undercooling of {\it l}-Si in the NPH ensemble, it is possible to show
unequivocally that the system undergoes a first order liquid-liquid
transition from a high density liquid (HDL) to a low density liquid
(LDL) before transforming continuously to the amorphous state, a
thermodynamically metastable state.

\begin{figure}
\centering
\includegraphics[width=2.4in,angle=-90]{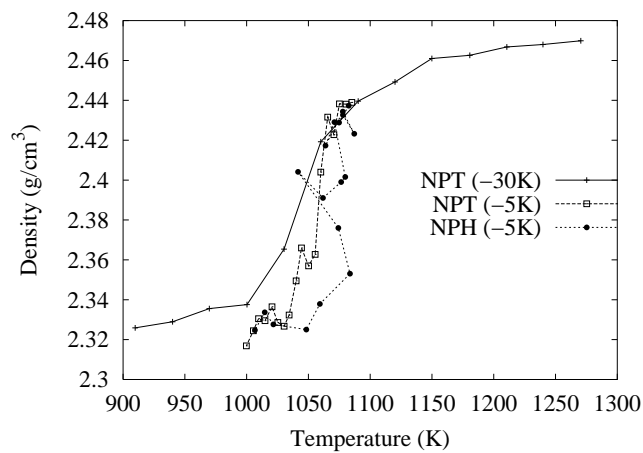} 
\caption{
  Mean density as a function of temperature for a 1000-atom SW cell at
  zero pressure. Simulations start with a well-equilibrated model of
  {\it l}-Si at 1300 and 1090 K for temperature steps of -30 and -5 K
  respectively. The first two simulations are performed in the NPT
  ensemble, with the temperature lowered by steps of 30 K (crosses and
  solid line) and 5 K (open squares and dashed line) while the last
  curve shows the density as a function of temperature for a
  simulation in the NPH ensemble in which the temperature is decreased
  by steps of 5 K (filled circles and dotted line). Lines are guide to
  the eye.  }
\label{thermo_P0_-5K}
\end{figure}

Our simulations find the same first-order liquid-liquid phase
transition near 1060 K as the system is cooled slowly from high
temperature. Figure~\ref{thermo_P0_-5K} shows the density as a
function of the temperature in the NPT ensemble with steps of -30 and
-5 K between each point while the NPH data is obtained with steps of
-5 K. Small temperature steps are needed to notice the details of the
first-order liquid-liquid transition. Each of these points is obtained
by averaging over 450 000 time steps, after a 50 000 time step
equilibration period.  As explained in Ref.~\cite{Sastry}, the
discontinuities in the NPH curve are associated with a first-order
transition.
The transition is also clearly visible in the density: the system
expands brutally as it is cooled from 1090 to 1050 K and its density
falls from 2.44 to 2.32 g/cm$^3$. This change in volume is associated
with a lowering of coordination from 4.9 to 4.24 as the system becomes
more like a network liquid.

\begin{figure}
\centering
\includegraphics[width=4.0in]{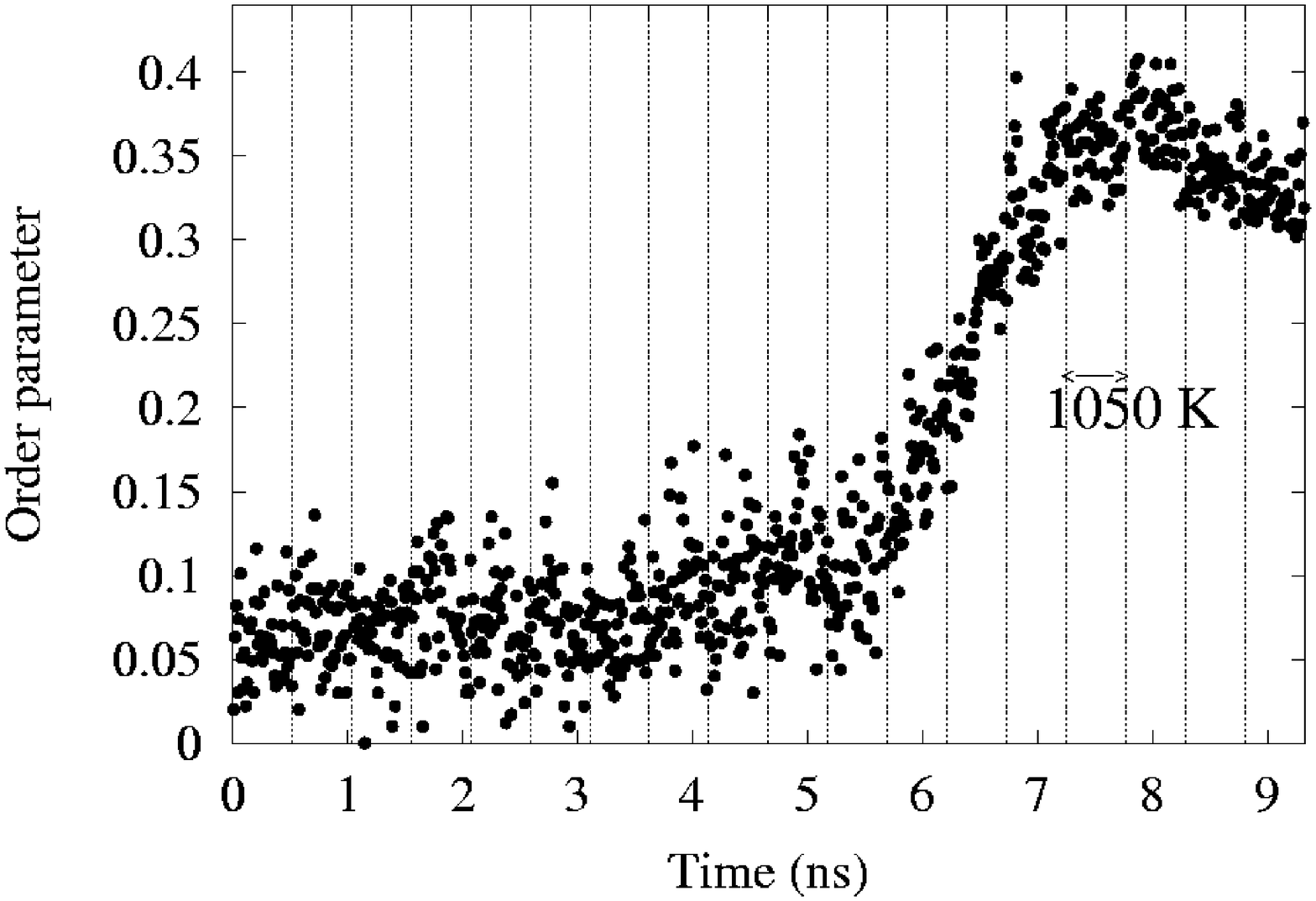} 
\caption{
  Evolution of the proportion of atoms in crystalline structures for
  the simulation in the NPH ensemble at zero pressure.  Each section
  represents a time interval at a given enthalpy with the kinetic
  energy reduced by 5 K at the end of each interval.  Each interval
  correspond to a thermodynamic point on the density vs temperature
  figure (see Fig. ~\ref{thermo_P0_-5K}). The double arrow indicated
  time interval for which the average temperature is 1050 K.}
\label{crit_P0_-5K}
\end{figure}

The LDL possesses a largely tetrahedral structure with a coordination
of 4.24 at 1050 K in NPH ensemble, near the 4.2 value reported by
Sastry and Angell \cite{Sastry}.  The topological analysis, using the
elementary blocks, shows that 37 \% of the atoms are associated with a
basic building block; the structure is partially crystallized (Fig.
\ref{crit_P0_-5K}) whereas the HDL phase at T = 1090 K contains only
5-10 at.\% of elementary blocks.  The diffusion of atoms changes
abruptly during the transition and the diffusivity of the quenched
liquids at 1050 K is 3.2 $\cdot 10^{-8}$ and 3.8 $\cdot 10^{-7}$
cm$^2$/s in NPH and NPT ensembles, respectively, corresponding to a
very viscous liquid.

\begin{figure}
  \centering
  \includegraphics[width=3in,angle=-90]{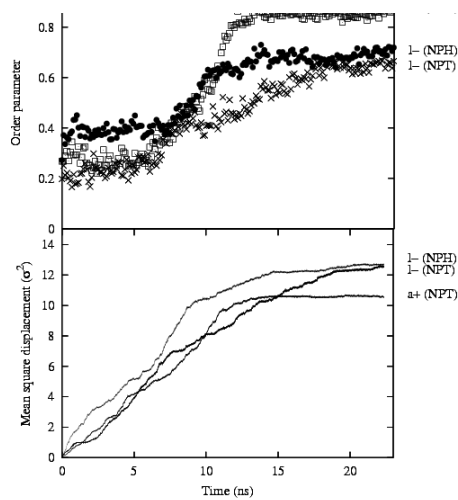}
\caption{Proportion of atoms belonging to basic building-block
  structures (top) and mean square displacement per atoms (bottom) for
  simulation performed at a constant temperature (T = 1050 K). The
  initial liquid configurations at 1050 K are obtained from the
  cooling simulations in the NPT "l- (NPT)" and NPH "l- (NPH)"
  ensemble and are run in the same ensemble. The initial amorphous
  model was obtained by heating to 1050 K in the NPT ensemble "a+
  (NPT)".}
\label{diff_T1050K}
\end{figure}

We verify that the LDL phase is the liquid counterpart to amorphous
silicon. To check this, we heated slowly a high-quality amorphous
model generated by a bond-switching technique described in
Ref.~\cite{Barkema00} and relaxed with the SW potential, and compared
with the LDL system. The amorphous phase is stable and does not
crystallize after a long simulation (20 ns) at 1000 K. Its diffusivity
increases rapidly at 1050K, however, and it reaches a value very
similar to that of the LDL at the same temperature, crystallizing
within 12 ns (with 80 at.\% of elementary blocks). This nucleation
time is similar to that observed in the LDL obtained by cooling in the
NPT and NPH ensemble (see Fig.\ ~\ref{diff_T1050K}).  The degree of
crystallization in Fig.\ ~\ref{diff_T1050K} (top) after 20 ns seems to
be closely related to the quality of the network in the initial state,
as defined by its coordination. The amorphous model, with an initial
coordination 4, crystallizes to 84\% while only 66 and 71 \% of atoms
belong to basic block structures for the liquids cooled in the NPT and
NPH ensemble, with an average coordination of 4.41 and 4.24 respectively.

Since the phase transition is buried deeply into the supercooled
region, it is difficult to reverse the transition and go from LDL to
HDL. Heated by step from 1050 K, the LDL phase remains stable until it
crystallizes at around 1150 K, showing considerable hysteresis. In
order to reach HDL, it is necessary to bring the LDL system at once
from 1050 to 1250 K, a behavior already noticed by Broughton and
Li~\cite{Broughton}.  The HDL phase, on the contrary, is stable
against crystallization and does not show any significant density of
elementary crystalline blocks even after 20 ns of simulation at a
temperature as low as 1100 K.

\subsubsection{Negative pressure}
\label{P_neg}

As discussed in the previous section, the LDL phase of the SW
potential shows an average coordination of 4.24 at 1050 K, indicating
the presence of a large fraction of mostly 5-fold coordinated atoms.
This coordination is significantly higher than the experimental
measurement, which showss an average coordination of
3.88~\cite{Laaziri}, and than the theoretically accepted value of
4.0~\cite{Barkema96, Vink, Barkema00}.  Such discrepancy is largely
due to the limits of the potential.  As discussed in
Ref.~\cite{Barkema96}, for example, the SW potential systematically
produces an over-coordinated amorphous phase; even very slow cooling
fails to produce configurations near an average coordination of
4.0~\cite{Luedtke2}.

Because of this discrepancy, it is important to test the stability of
the LDL phase as the system is biased towards obtaining a
higher-quality amorphous phase.  It is possible to favor the formation
of a lower-coordinated liquid by applying a negative pressure on the
system or by changing the potential. In this section, we study the HDL
to LDL at a negative pressure of -2 GPa, near to the stability limit
of the liquid; effects of a potential change are studied in the next
section.

\begin{figure}
\centering
\includegraphics[width=2.5in,angle=-90]{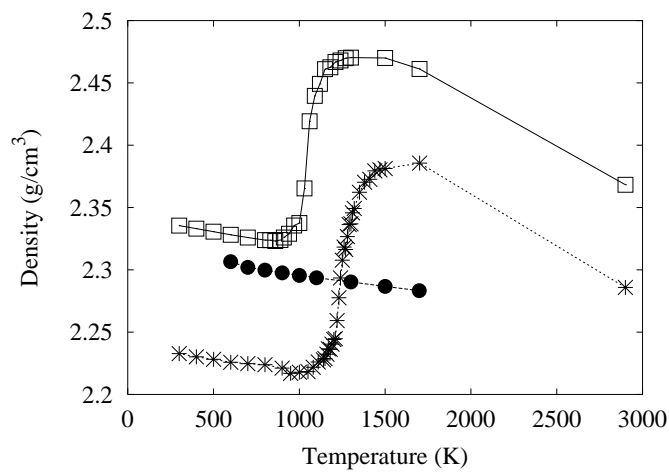} 
\caption{Density dependence of temperature for liquid and crystalline
  silicon. We show the results for the crystal ($\bullet$) and liquid
  at zero GPa ($\square$) and -2 GPa ($\ast$). At zero pressure, the
  supercooled liquid undergoes a first order transition at around 1060
  K. Lines are a guide to the eyes.}
\label{densite01}
\end{figure}

To first order, the application of negative pressure simply shifts the
phase diagram (Fig.~\ref{densite01}), as the density maximum moves
from 1300 K to 1700 K and the liquid-liquid transition from 1060 K to
1250 K, in agreement with the isochoric cooling curves as a function
of density presented by Angell {\it et al.}~\cite{Angell}.

\begin{figure}
\centering
\includegraphics[height=4.0in,angle=-90]{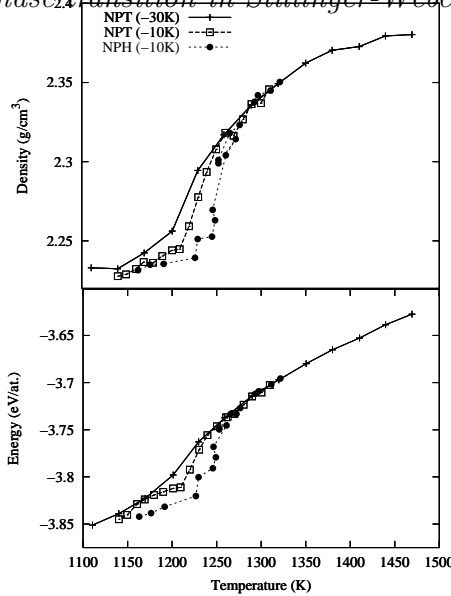} 
\caption{Mean density (top) and energy (bottom) as a function
  of temperature for a system maintained at a negative pressure of -2
  GPa.  The crosses (solid line) show the results of a NPT simulation
  starting from a well-relaxed liquid configuration at 1500 K and then
  gradually lowered by steps of 30 K. The open squares (dashed line)
  and the filled circles (dotted line) correspond, respectively, to
  simulations in the NPT and NPH ensemble with an initial liquid
  configuration equilibrated at 1340 K where the temperature is
  lowered by steps of 10 K. Lines are guide to the eye.}
\label{thermo_P-02_-10K}
\end{figure}

Similarly, the HDL to LDL transition remains at negative pressure.
However, the system does not emit latent heat during the transition
and NPH simulations, with cooling steps of 10 K, closely follow the
NPT curve (see Fig. ~\ref{thermo_P-02_-10K}). The HDL transforms,
therefore, via a second-order phase transition into a lower-density
phase. During the transition, between 1270 and 1225 K, the system
transforms in a tetravalent structure with the average coordination
falling from 4.95 to 4.18 in NPH conditions and the proportion of
atoms in crystalline structures increases from 5 \% to 30 \%. At 1225
K, the diffusivity of the LDL is $1.6 \cdot 10^{-7}$ cm$^2$/s, and
is therefore still a liquid.

\begin{figure}
  \centering
\includegraphics[width=4.0in]{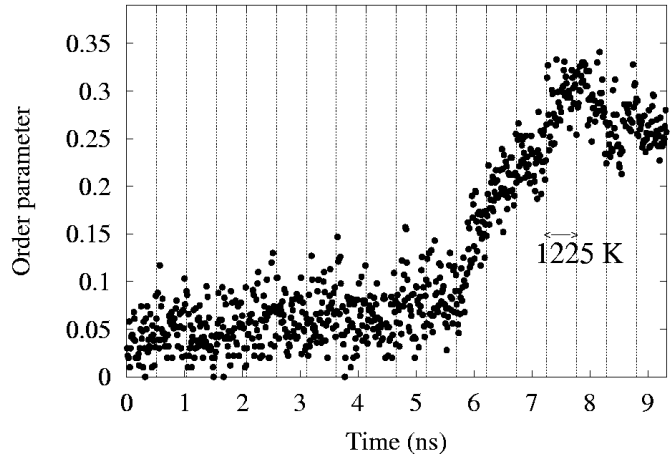} 
\caption{Evolution of the proportion of atoms in crystalline
  structures for the simulation in NPH ensemble at a negative pressure
  of -2 GPa.  Each section represents a time interval at a given
  enthalpy with the kinetic energy reduced by 5 K at the end of each
  interval.  Each interval correspond to a thermodynamic point on the
  density vs temperature figure (see Fig. ~\ref{thermo_P-02_-10K}).
  The double arrow indicated time interval for which the average
  temperature is 1225 K.}
\label{crit_P-02_-10K}
\end{figure}

The theoretical model of Aptekar~\cite{Aptekar} locates the
liquid-liquid transition at around 1500 K for Si at zero pressure and
predicts a second critical point that terminates the first-order
transition at -1.5 GPa in the supercooled liquid region. At -2 GPa,
this model should therefore be below the critical point. Our picture
is more complex as there is now a second-order phase transition
between the HDL and LDL, indicating that critical point does not
correspond to the end of a coexistence line.

\subsubsection{Modified potential}
\label{mmSW}


High-quality amorphous structures require a significant strengthening
of the three-body term of the SW potential; an increase of 50 \% is
necessary to ensure a coordination of 4 in the amorphous phase for
this potential~\cite{Vink, Luedtke, Broughton}. This strong
modification of the potential, however, completely changes the phase
diagram of the system, removing the maximum in the density of the
liquid phase and setting a lower density to the liquid than the
crystal. Although a 12.5 \% modification to the three-body term
restores a temperature of density maximum in the liquid phase, the
amorphous (or glassy) state reaches a lower density than the crystal.
Thus, in order to preserve the main features of the SW phase diagram,
we use much weaker modification, increasing the three-body term by 5
\%, in order to verify the universality of the liquid-liquid phase
transition. 

\begin{figure}
\centering
\includegraphics[width=2.5in,angle=-90]{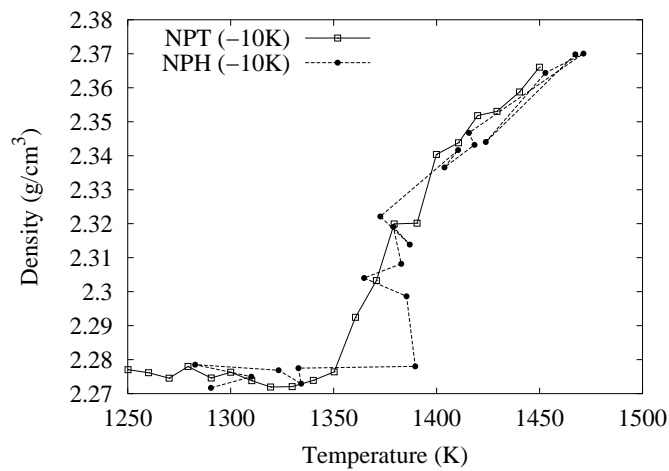} 
\caption{Density as a function of temperature
  with the modified SW potential ($ \lambda = 1.05 \lambda_0 $). The
  open squares (solid line) and the filled circles (dashed line) show
  the result of simulation in the NPT and NPH ensemble,
  respectively, with cooling by steps of 10 K starting from an
  equilibrated {\it l}-Si configuration at 1460 K. Lines
  are guide to the eye.}
\label{thermo_m1.05SW}
\end{figure}

The transition from the high to low density phase has features similar
to that for the original SW potential. In particular, a net release of
latent heat confirms that the transition is first order. However, the
small strengthening of the three-body term augments significantly the nucleation
rate of Si and  raises the transition temperature from 1060 K to 1390
K (Fig.  ~\ref{thermo_m1.05SW}), indicating that the transition is
qualitatively different.  Indeed, the new phase, obtained following
either the NPT or the NPH conditions, is already considerably
crystalline at the transition: local-order analysis shows that the NPH
configuration at 1390 K and 2.28 g/cm$^3$ occuring during the
transition is crystallizing, with 55 \% of all atoms belonging to
crystalline structures.
The liquid-liquid transition is therefore totally hidden by the
crystallization. 

\begin{figure}
\centering
\includegraphics[width=4.0in]{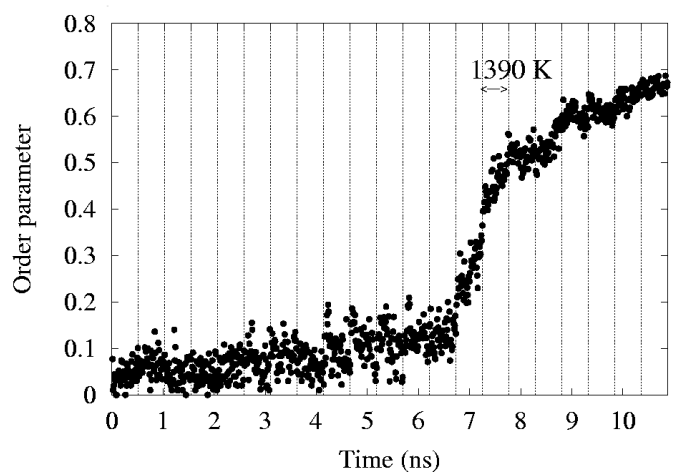} 

\caption{Evolution of the proportion of atoms in crystalline
  structures for a simulation in the NPH ensemble with the modified SW
  potential.  Each section represents a time interval at a given
  enthalpy with the kinetic energy reduced by 5 K at the end of each
  interval.  Each interval correspond to a thermodynamic point on the
  density vs temperature figure (see Fig. ~\ref{thermo_m1.05SW}). The
  double arrow indicated time interval for which the average
  temperature is 1390 K.}
\label{thermo_m1.05SW}
\end{figure}

\begin{figure}
\centering
\includegraphics[width =2.5in,angle=-90]{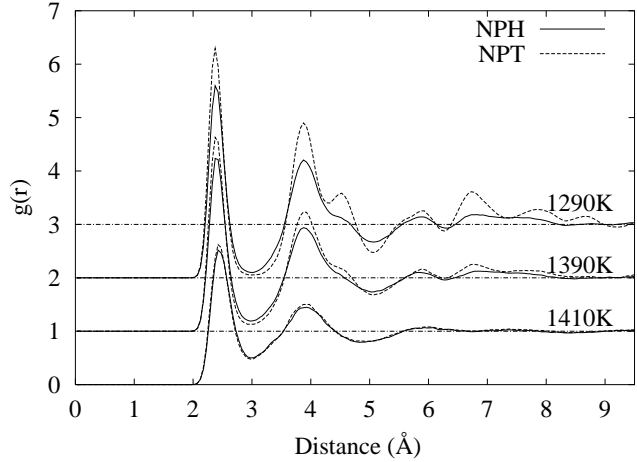} 
\caption{Radial distribution function of the modified SW potential
  near the transition temperature in the NPT and NPH ensemble. The RDF
  are measure at different temperatures: 1290 K (top), 1390 (middle)
  and 1410 K (bottom).  }
\label{rdf_m1.05SW}
\end{figure}

Because of the crystallization, the NPH and NPT simulations do not
finish in the same thermodynamical state after the transition,
contrarily to what is observed with the original SW potential: the NPT
simulation shows a much more rapid crystallization than the NPH and at
1290 K, more than 95 \% of all atoms are in a crystalline environment
against 67 \% for the NPH simulation. This high degree of
crystallinity can be seen directly in the RDF, as shown in Fig.
~\ref{rdf_m1.05SW}: at 1390 K, only the NPT simulation shows the
third-neighbor peak, while it is visible also in the NPH simulation at
1290 K.

\section{Discussion and conclusion}
\label{discu}

The liquid-liquid transition in Si is difficult to observe because it
should occur deep in the undercooled region of the phase diagram. It
is not clear whether this part of the phase diagram can be reached
experimentally, so we must rely on simulations to characterize this
important phenomenon. Even with the fast cooling accessible to
molecular dynamics, the temperature window for observing the LDL is
very narrow and the viscosity increases rapidly as the LDL becomes
glassy {\it a}-Si. This difficulty explains why the phase was only
demonstrated recently for SW Si. 

At zero pressure, Sastry and Angell~\cite{Sastry} demonstrated clearly
the existence of a liquid-liquid phase transition in the SW Si.
Moreover, we could verify that the amorphous phase, obtained via an
independent method, does correspond to the LDL, as was suggested
previously.

The question remains as to whether the LDL phase also exists in the
real material. By moving around this thermodynamical point, it is
possible to verify the stability of this result and whether or not
this phase is likely to occur in Si. 

As one of the main limitations of the SW potential is that it cannot
describe properly the structure of the amorphous phase, corresponding
to the glassy phase of LDL, we look at the impact of getting a better
coordinated {\it a}-Si on the phase diagram by applying a negative
pressure and changing slightly the potential.

Following the analysis of Aptekar, a critical point should exist at
the end of the coexistence line below -1.5 GPa~\cite{Aptekar}. We find
instead that the order of the liquid-liquid transition changes from
first to second, with an absence of heat release during the
transition.  The relative change in structure from HDL to LDL is very
similar to that obtained at zero pressure, however, and the average
coordination is closer to 4.0. Nevertheless, the LDL seems to be even
more unstable under crystallization than at zero pressure. 

It is also possible to favor a lower coordination in the amorphous
phase by increasing the strength of the three-body force of the SW
potential. To keep the same overall phase diagram, we modify this term
only very slightly, increasing it by only 5 \%. The impact of this
modification is surprisingly important: the first-order transition
move from 1060 to 1390 K and changes in character, crystallization
occurs almost immediately and there is no trace of a low-density
liquid. The temperature shift is much larger than the change in
melting temperature, which is about 20 K, suggesting that the
transition seen with the strengthened potential is the standard
crystallization transition in the undercooled phase. The liquid-liquid
transition, if present, is therefore not easily reachable even on MD
timescale. This is particularly clear using a topological
order-parameter which identifies the degree of crystallinity with much
more precision than averaged structural quantities such as the
structure factor and the RDF.

While the results of Sastry and Angell provided the first clear
demonstration of the existence of a liquid-liquid first-order
transition in SW Si~\cite{Sastry}, our results suggest that the
existence and the nature of the liquid-liquid phase transition in real
Si must be confirmed through further simulations with a wider set of
potentials and, if possible, through experiment.

\ack
This work is funded in part by NSERC, NATEQ and the Canada Research
Chair Program.  NM is a Cottrell Scholar of the Research Corporation.
Most of the simulations were run on the computers of the R\'eseau
qu\'eb\'ecois de calcul de haute performance (RQCHP) whose support is
gratefully acknowledged.

\end{document}